\documentclass[12pt]{article}
\large

\title{Paths with singularities in functional integrals of quantum field theory.}

\vspace{0.5cm}

\author{ V.V. Belokurov and E.T. Shavgulidze    \\
{\em Lomonosov Moscow State University, Russia }
\\ {\it e-mail: belokur@rector.msu.ru}}

\date{ \ \ \  }
\begin{document}
\maketitle

\vspace{0.5cm}

The quantum field theory model studied  in our recent paper \cite{(BSh)} demonstrates that quantum properties of a model depend in a great extent on the functional space we integrate over.

Here we study a toy model: $ \varphi^{4}$-interacting quantum field theory in one-dimensional "Euclidean" space-time. We prove that the functional integrals of the free field theory evaluated over the space of continuous functions  are equal to the functional integrals of the interacting field theory evaluated over a set of spaces containing the spaces of discontinuous functions.

Consider the functional integral
\begin{equation}
   \label{1}
\int\,\exp\left\{ -\frac{1}{2}\int \limits _{0}^{1}(\dot{\varphi}(t))^{2}dt -\frac{1}{2}\int \limits _{0}^{1}\varphi^{4}(t)dt \right\}\ d\varphi\,.
\end{equation}
Here we suppose for simplicity that the field function $\varphi(t)$ (the stochastic process) is defined on the finite closed interval $[0,1]$.

The integral (\ref{1}) exists for continuous field functions $(\,\varphi(t)\in C[0,1]\,)\,.$ It is the integral over the Wiener measure $\exp\{ -\frac{1}{2}\int \limits _{0}^{1}(\dot{\varphi}(t))^{2}dt \}\ d\varphi\,,$ the integrand $\exp\{-\frac{1}{2}\int \limits _{0}^{1}\varphi^{4}(t)dt\}$ being a bounded functional.

Notice that here all the derivatives are understood in a generalized sense.

After the substitution
\begin{equation}
   \label{2}
\chi(t)=\varphi(t)+\int \limits _{0}^{t}\varphi^{2}(\tau)d\tau\,,
\end{equation}
formally we get
$$
\dot{\chi}(t)=\dot{\varphi} (t)+\varphi^{2}(t)\,,
$$
and
\begin{equation}
   \label{3}
\frac{1}{2}\int \limits _{0}^{1}(\dot{\chi}(t))^{2}dt=\frac{1}{2}\int \limits _{0}^{1}(\dot{\varphi}(t))^{2}dt +\frac{1}{2}\int \limits _{0}^{1}\varphi^{4}(t)dt+
\int \limits _{0}^{1}\dot{\varphi}(t)\varphi^{2}(t)dt\,.
\end{equation}

The last term in the equation (\ref{3}) is the Ito stochastic integral \cite{(Ito)}
$$
\int \limits _{(t=0)}^{(t=1)}\varphi^{2}\,d\varphi = \frac{1}{3}\left[\varphi^{3}(1)-\varphi^{3}(0)\right]-\int \limits _{0}^{1}\varphi(t)dt\,.
$$

So, we can make the conclusion about the formal equality of the integrals
$$
\int\,\exp\left\{ -\frac{1}{2}\int \limits _{0}^{1}(\dot{\varphi}(t))^{2}dt -\frac{1}{2}\int \limits _{0}^{1}\varphi^{4}(t)dt- \frac{1}{3}\left[\varphi^{3}(1)-\varphi^{3}(0)\right]+\int \limits _{0}^{1}\varphi(t)dt\right\}\ d\varphi
$$

and
$$
\int\,\exp\left\{ -\frac{1}{2}\int \limits _{0}^{1}(\dot{\chi}(t))^{2}dt \right\}\ d\chi \,.
$$
It can be easily verified using the discrete versions of the integrals where the substitution (\ref{2}) looks like
$
\chi(t_{k})=\varphi(t_{k})+\frac{1}{N}\sum_{i=o}^{k-1}\varphi^{2}(t_{i})\,.
$

 So far, we have not specified the functional spaces we integrate over. As we show in this letter, for the equality of these integrals to be valid the functional spaces should be different. Namely,
$$
 \int\limits _{C[0,1]}\,\exp\left\{ -\frac{1}{2}\int \limits _{0}^{1}(\dot{\varphi}(t))^{2}dt -\frac{1}{2}\int \limits _{0}^{1}\varphi^{4}(t)dt - \frac{1}{3}\left[\varphi^{3}(1)-\varphi^{3}(0)\right]+\int \limits _{0}^{1}\varphi(t)dt\right\}\ d\varphi\,=
$$
\begin{equation}
   \label{4}
\int\limits _{Y}\,\exp\left\{ -\frac{1}{2}\int \limits _{0}^{1}(\dot{\chi}(t))^{2}dt \right\}\ d\chi \,\neq\,1\,,
\end{equation}
where
$
Y\subset\,C[0,1]\,,\ \ \  Y\neq\,C[0,1]\,.
$
Conversely,
$$
1\,=\,\int\limits _{C[0,1]}\,\exp\left\{ -\frac{1}{2}\int \limits _{0}^{1}(\dot{\chi}(t))^{2}dt \right\}\ d\chi \,=
$$
\begin{equation}
   \label{5}
\int\limits _{X}\,\exp\left\{ -\frac{1}{2}\int \limits _{0}^{1}(\dot{\varphi}(t))^{2}dt -\frac{1}{2}\int \limits _{0}^{1}\varphi^{4}(t)dt- \frac{1}{3}\left[\varphi^{3}(1)-\varphi^{3}(0)\right]+\int \limits _{0}^{1}\varphi(t)dt\right\}\ d\varphi\,,
\end{equation}
where
$
X\neq\,C[0,1]\,,\ \ \ X\supset\,C[0,1]\,.
$

Now, let us study the structure of the space $X$.
Consider a function $\chi(t)\in C[0,1]$. Let us find the general form of the function $\varphi(t)$ satisfying the equation (\ref{2}).
For the function $\mu(t)=\chi(t)-\varphi(t)$ we have
$$
\mu(t)=\int \limits _{0}^{t}\left(\mu(\tau)-\chi(\tau) \right)^{2}d\tau\,,
$$
and
$$
\dot{\mu}(t)=\left(\mu(t)-\chi(t) \right)^{2}\,.
$$
So, there is a point  $t_{1}$ where the function $ \mu(t_{1})>0$.
  For the function
  $$
  \nu (t)=\frac{1}{\mu(t)}
  $$
 the differential equation looks like
$$
\dot{\nu}(t)=-(1-\chi(t)\nu(t))^{2}\,.
$$
As $\dot{\nu}(t)<0$ and $\nu(t_{1})>0$ there can be a point $t^{\ast}\in[t_{1},1]$ where $\nu(t^{\ast})=0$.
In the vicinity of this point we have
$$
\nu (t)=-\int \limits _{t^{\ast}}^{t}\,\left(1-\chi(\tau)\nu(\tau)\right)^{2}d\tau=-(t-t^{\ast})-\chi(\tilde{t})(t-t^{\ast})^{2}+O((t-t^{\ast})^{3})\,,\ \ \ \tilde{t}\in[t^{\ast},t]\,,
$$
and
$$
\mu(t)=- \frac{1}{t-t^{\ast}}+\chi(\tilde{t})+O((t-t^{\ast}))\,,\ \ \ \tilde{t}\in[t^{\ast},t]\,.
$$
That is, the function $\varphi(t)$ has a singularity at $t=t^{\ast} $ and in the vicinity of the point the function $\varphi(t)$ has the form
$
\varphi(t)= (t-t^{\ast})^{-1}+ O((t-t^{\ast}))\,.
$

As the function $\chi(t)$ is  bounded on $[0,1]$, there is a finite interval $[t_{1},t_{2}]\subset [0,1]$ where
the function $\varphi(t)$ has the only singularity
$ (t-t^{\ast})^{-1}\,.$

Depending on the form of the function $\chi(t)$ there can be other finite intervals $[t_{3},t_{4}]\,,\ldots$ where $\varphi(t)$ has singularities of the same type $ (t-t_{j}^{\ast})^{-1}$ and the behavior of the function  $\varphi(t)$ in the vicinity of the point  $t_{j}^{\ast}$ is given by the equation
\begin{equation}
   \label{6}
\varphi(t)= \frac{1}{t-t_{j}^{\ast}}+ O((t-t_{j}^{\ast}))\,.
\end{equation}

Due to the compactness of the interval $[0,1]$  the number of the singularities is finite $(j=1,\ldots, n\,)$ for every $\chi(t) \,.$

To elucidate the result obtained we notice that for the bounded function $\chi(t)$ there can be some regions where $|\varphi(t)|$ becomes large enough. And in these regions the behavior of the function $\varphi(t)$ is prescribed by the equation $\dot{\varphi}=-\varphi^{2}\,.$

Now, the space $X$ can be represented in the form
$$
X=X_{0}\cup X_{1}\cup X_{2}\cup\ldots X_{n}\cup\ldots\,,
$$
where $X_{0}=C[0,1]$ and $X_{n}$ is the space of functions with $n$ singularities of the type (\ref{6}).

The equation (\ref{5}) looks like
$$
1\,=\,\int\limits _{C[0,1]}\,\exp\left\{ -\frac{1}{2}\int \limits _{0}^{1}(\dot{\chi}(t))^{2}dt \right\}\ d\chi \,=
$$
\begin{equation}
   \label{7}
\sum_{n=o}^{\infty}\ \int\limits _{X_{n}}\,\exp\left\{ -\frac{1}{2}\int \limits _{0}^{1}(\dot{\varphi}(t))^{2}dt -\frac{1}{2}\int \limits _{0}^{1}\varphi^{4}(t)dt- \frac{1}{3}\left[\varphi^{3}(1)-\varphi^{3}(0)\right]+\int \limits _{0}^{1}\varphi(t)dt\right\}\ d\varphi\,.
\end{equation}

We can easily evaluate  \cite{(Kuo)} the functional integral
$$
\int\limits _{C[0,1]}\,\exp\left\{ -\frac{1}{2}\int \limits _{0}^{1}(\dot{\chi}(t))^{2}dt +\imath \int \limits _{0}^{1}\chi(t)\eta (t)dt\right\}\ d\chi \,=
$$
\begin{equation}
   \label{8}
\exp\left\{ -\frac{1}{2}\int \limits _{0}^{1}\int \limits _{0}^{1}\min (t_{1},t_{2})\,\eta (t_{1})\eta (t_{2})\, dt_{1}dt_{2}\right\}\,.
\end{equation}
However, the functional integrals of the form
$$
\int\limits _{X_{n}}\,\exp\left\{ -\frac{1}{2}\int \limits _{0}^{1}(\dot{\varphi}(t))^{2}dt \right\}\,\ \emph{P}(\varphi)\,d\varphi\,,\ \ \ n\geq1\,,
$$
where $\emph{P}(\varphi)$ is a polynomial, do not exist.
$
\exp\left\{ -\frac{1}{2}\int \limits _{0}^{1}(\dot{\varphi}(t))^{2}dt \right\}\,d\varphi
$
is not a measure on $X_{n}\,,\ ( n\geq1)\,.$ (It forces us to recall the well-known Haag theorem (see e.g. \cite{(Str)}) ).
 And only
\begin{equation}
   \label{9}
  \exp\left\{ -\frac{1}{2}\int \limits _{0}^{1}(\dot{\varphi}(t))^{2}dt -\frac{1}{2}\int \limits _{0}^{1}\varphi^{4}(t)dt- \frac{1}{3}\left[\varphi^{3}(1)-\varphi^{3}(0)\right]+\int \limits _{0}^{1}\varphi(t)dt\right\}\ d\varphi
\end{equation}
can be considered as a measure on $X_{n}\,,\ (n\geq1)\,.$

We would like to stress that in the equation (\ref{2}) the functions $\chi(t)$ and $\varphi(t)$ are in one-to-one correspondence. However, it is not valid for the substitutions
$\chi(t)=\varphi(t)+\int \limits _{0}^{t}\varphi^{k}(\tau)d\tau$ with $k>2\,.$

In realistic quantum field theory (with the dimension of space-time $d=4$) the interacting fields are not the continuous functions but distributions. That is why, we have to consider functional integrals over the functional spaces more complicated than $C[0,1]\,.$


\begin{thebibliography}{4}

\bibitem {(BSh)}Belokurov V.V. and Shavgulidze E.T. On the local limit of quantum
field theories defined on the loop space. arXiv:1109.5954v1. [hep-th].

\bibitem {(Ito)} McKean H.P. Stochastic Integrals. Academic Press. New York - London. 1969.

\bibitem {(Kuo)} Kuo Hui-Hsiung. Gaussian measures in Banach spaces. Springer. Berlin-Heidelberg-New York. 1975.

\bibitem {(Str)} Streater R.F. and Wightman A.S. PCT, Spin and Statistics and All That. W.A. Benjamin, INC. New York - Amsterdam. 1964.

\end{thebibliography}
\end{document}